\documentstyle[prl,aps,twocolumn,epsf]{revtex}

\begin{document}
\draft
\title{
Instability due to  long range Coulomb interaction 
in a liquid of polarizable particles (polarons, etc.) 
}
\author{J. Lorenzana\cite{adr}}
\address{
Istituto Nazionale di Fisica della Materia - Dipartimento 
di Fisica, Universit\`a di Roma ``La Sapienza'',\\ Piazzale 
A. Moro 2, I-00185 Roma, Italy}

\date{\today}
\maketitle
\begin{abstract} 
The interaction Hamiltonian for a system of polarons {\it a la Feynman} in the presence of  
long range Coulomb  interaction is derived and the dielectric function
is  computed in mean field. For large enough
concentration a liquid of such particles becomes unstable. The onset of the instability is 
signaled by the softening of a collective optical mode in which all electrons 
oscillate in phase in their respective  self-trapping potential. We associate 
the instability with a metallization of the system. 
  Optical  experiments in slightly doped cuprates and doped nickelates 
are analyzed  within this theory. We discuss why doped cuprates matallize 
whereas nickelates do not.  
\end{abstract}
\pacs{Pacs Numbers:
71.10.-w, 
71.38.+i, 
78.20.Bh,
71.10.Hf 
}

\narrowtext
An electron moving in a highly polarizable lattice can distort 
the environment  and create a potential which self-traps the electron.
 The  resulting object can  move by dragging the distortion 
resulting in a heavy particle called polaron\cite{kit66}. 
Apart from its motion as a whole 
the polaron has an extra degree of freedom which can be thought as the 
electron oscillating in the self-trapping potential. 
Because of this internal mode a polaron can be polarized in an 
external electric field. 

 In this Letter a model Hamiltonian that takes into account
 the interactions due to mutual polarization is derived. It is shown  that a
 polaron 
liquid or more generally a liquid of polarizable particles  becomes
unstable for high enough concentrations due to the dipole-dipole
 interaction.   The onset of the instability is indicated by a 
softening\cite{lup99} of the internal mode of the polaron which can be 
measured in optical 
experiments\cite{lup99,xia93,xia93b,lup92,cal94,tho91,fal93}. 
Recent experimental results in  cuprates and 
nickelates are analyzed within this scenario\cite{lup99,xia93,xia93b}. 

For simplicity the present treatment is semiclassical (in the spirit of the 
Drude model\cite{kit66}) however we expect that the same effect will show up
in a fully  quantum mechanical computation.

We  concentrate on the case of dielectric polarons
but we believe our results are valid also for other kinds of heavily
dressed particles in solids as long as they have a polarizable
internal degree of freedom.

Consider an electron in a dielectric with  static 
dielectric constant $\epsilon_0$ dominated by {\em lattice polarization}. 
 If the electron is not allowed to move
it will generate a radial polarization of the lattice towards itself.  This 
distortion implies that some ionic positive  charge ($q$)  will
accumulate in the vicinity of the electron.  The charge induced in the 
lattice can be estimated as
\begin{equation}
  \label{eq:q}
  q=e\left( 1-\frac1{\epsilon_0} \right).
\end{equation}
Now let us free the electron. If the electron binds to the induced lattice 
charge  we have a dielectric polaron. In this case for the 
electron to move it has to drag  the distortion of the lattice.

To describe this situation we use  Feynman's effective model for the
polaron\cite{fey55,fey70}. The effect of the distortion is mimic by a heavy
fictitious particle of mass $M$. The electron, of mass $m$, is coupled 
to the heavy particle   with a harmonic potential. The result is a 
composite particle. The non interacting  Hamiltonian for  a liquid of  such
 particles reads $H_0=\sum_i H_i$ with
\begin{equation}
H_i=\frac{p_i^2}{2m}+\frac{P_i^2}{2M}+\frac{k}2 |\bbox{r}_i-\bbox{R}_i|^2. 
 \label{eq:hi}
\end{equation}
The first term in $H_i$ is the kinetic energy of the electron. 
The second term is the kinetic energy of the heavy mass representing the 
surrounding deformation.  The third term is the coupling
between the electron and the distortion. $\bbox{p}_i$, $\bbox{r}_i$ 
($\bbox{P}_i$, $\bbox{R}_i$)
are the momentum and position of the electron (fictitious mass) $i$. 

In the original formulation by Feynman the action corresponding to 
 Eq.~($\ref{eq:hi}$)  is regarded as an effective action 
for the dynamics of a polaron. The 
parameters are obtained variationally from the Fr\"ohlich
 model\cite{fro54} of a  dielectric  polaron. Hare we take as granted that 
$H_0$ is a first approximation
to describe the  dynamics  of a polaron liquid and explore the 
consequences of adding a Coulomb term. 
 On physical grounds we will 
assume the charge $q$ is distributed with spherical symmetry  centered 
on $R_i$.

 To derive the interaction Hamiltonian consider two fixed electrons in the 
dielectric at $\bbox{r}_1,\bbox{r}_2$. 
As before each electron will induce a close-by charge in the 
lattice of  magnitude  $+q$. The total electrostatic energy can be 
computed as usual. 
One gets self-energy terms plus the interaction term: 
\begin{equation}
  \label{eq:int}
  E_{\rm int}= \frac{e^2}{\epsilon_0|\bbox{r}_1-\bbox{r}_2|}.
\end{equation}
This energy includes the vacuum electrostatic energy between  charges 
plus the elastic energy stored in the dielectric:
\begin{equation}
  \label{eq:eint}
  E_{\rm int}=E_{\rm Coul}+E_{\rm ela}.
\end{equation}
The former is:
\begin{equation}
  \label{eq:cou}
  E_{\rm Coul}= \frac{e^2}{\epsilon_0^2|\bbox{r}_1-\bbox{r}_2|}
\end{equation}
i.e. the electrostatic energy between the screened charges. 
 Eq.~(\ref{eq:eint}) can be solved for the elastic energy:
\begin{equation}
  \label{eq:ela}
  E_{\rm ela}= \frac{e^2}{|\bbox{r}_1-\bbox{r}_2|}\frac1{\epsilon_0} 
\left(1-\frac1{\epsilon_0}\right).
\end{equation}
The factors with the dielectric constant in Eq.~(\ref{eq:ela}) ensure 
that the elastic energy is zero for an infinitely rigid lattice
($\epsilon_0\rightarrow 1$) or an infinitely soft lattice 
($\epsilon_0\rightarrow \infty$).

Suppose now that we move the electrons  from the equilibrium
position  keeping  ions fixed so that the distortion does not move.
i.e.  $\bbox{r}_i\ne \bbox{R}_i$. 
Since the elastic energy depends only on the configuration of the
lattice we should replace in Eq.~(\ref{eq:ela}) 
$\bbox{r}_i\rightarrow \bbox{R}_i$. 
So that in general the elastic energy is:
\begin{equation}
  \label{eq:elag}
  E_{\rm ela}= \frac{(e-q)q}{|\bbox{R}_1-\bbox{R}_2|}
\end{equation}
where we eliminated $\epsilon_0$ using the definition of the induced lattice
charge $q$ [Eq.~(\ref{eq:q})].

The Coulomb energy can be decomposed into elementary Coulomb 
interactions. For general position of charges and distortions
it  reads:
\begin{equation}
  \label{eq:coug}
  E_{\rm Coul}= \frac{e^2}{|\bbox{r}_1-\bbox{r}_2|}
-\frac{ e q}{|\bbox{r}_1-\bbox{R}_2|} 
-\frac{ e q}{|\bbox{R}_1-\bbox{r}_2|}+ 
\frac{q^2}{|\bbox{R}_1-\bbox{R}_2|}
\end{equation}
which reduces to Eq.~(\ref{eq:cou})  for  $\bbox{r}_i=\bbox{R}_i$.

Adding again the 
elastic and Coulomb energy we can write the interaction Hamiltonian for a
liquid of such particles:
\begin{equation}
  \label{eq:hin}
  H_{\rm int}=\sum_{ij,i\ne j} 
  \frac12  \frac{e^2}{|\bbox{r}_i-\bbox{r}_j|} - \frac{eq}{|\bbox{r}_i-\bbox{R}_j|} 
+\frac12  \frac{eq}{|\bbox{R}_i-\bbox{R}_j| }  
\end{equation}
where indexes $i,j$ run over  particles.

A similar argument can be used to derive the interaction Hamiltonian
with an external electric field.
The Coulomb contribution of the induced charge cancels with the elastic part
and one obtains,
\begin{equation}
  \label{eq:HE}
  H_E=  \sum_i e \bbox{r}_i.\bbox{E}.
\end{equation}
The many-particle Hamiltonian is $H=H_0+H_{\rm int}+H_E $. 

  It is convenient to change to center of mass variables,
 $ \bbox{\rho }_i=(\bbox{r}_i m + \bbox{R}_i M)/(M+m)$, 
  $\bbox{u }_i=\bbox{r}_i-\bbox{R}_i$. Making a Taylor expansion for small  
$\bbox{u}_i$ we obtain  the following interacting Hamiltonian in the dipole 
approximation,
\begin{eqnarray}
  \label{eq:hinld}
    H_{\rm int}= \sum_{ij,i\ne j} &&
\frac12  \frac{e(e-q)}{|\bbox{\rho}_i-\bbox{\rho}_j| }
     -   \frac{ e (e - q)}{1+m/M} 
\frac{\bbox{u}_i.(\bbox{\rho}_i-\bbox{\rho}_j)}{|\bbox{\rho}_i-\bbox{\rho}_j|^3}\nonumber\\
-&& \frac{e (e-q)}{2(1+m/M)^2} \bbox{u}_i. \bbox{\phi}(\bbox{\rho}_i-\bbox{\rho}_j). \bbox{u}_i\\
+&& \frac12 
\left(\frac{e(e-q)}{(1+m/M)^2}+e q\right)\bbox{u}_i.\bbox{\phi}(\bbox{\rho}_i-\bbox{\rho}_j). \bbox{u}_j
\nonumber
\end{eqnarray} 
where we defined the matrix
 $ \phi_{\mu\nu}(\bbox{\rho})=\delta_{\mu\nu}/{\rho^3}-3\rho_\mu \rho_\nu/{\rho^5}$
and  $\mu,\nu$ are Cartesian indexes.

Hamiltonian Eq.~(\ref{eq:int}) have the essential
ingredients to treat the interacting polaron problem at concentrations 
such that the interparticle distance exceeds the polaron radius. 

Now we compute the dielectric constant of such a system.
To obtain  equations of motion for one particle in the mean field 
of the others we compute  the forces  
$\bbox{F}_{\bbox{u}_i}=-\partial H/\partial \bbox{u}_i$ and
$\bbox{F}_{\bbox{\rho}_i}=-\partial H/\partial \bbox{\rho}_i$. The former 
is the force that polarizes  particles and the latter is the force 
acting on the center of mass. The equations of motion in mean field read,
\begin{eqnarray}
  \label{eq:moti}
\mu \bbox{\ddot u} & = & -k\bbox{ u} +  4 \pi L n e q \bbox{ u} - 
\frac{ e} {1+m/M}\bbox{ E}\\
  (m+M)\bbox{\ddot  \rho}  &=& - e\bbox{ E}
\end{eqnarray}
where  $\bbox{u} \equiv<\bbox{u}_i>$, etc. $n$ is the density of particles,
$1/\mu=1/m+1/M$, and $L$ is a geometric factor discussed below. 

A crucial point in the derivation is the evaluation of the force due 
to  dipole-dipole interactions 
$\bbox{ F_L}=
e q  \bbox{u} . <\sum_{j,j\ne i}  \bbox{\phi}(\bbox{\rho}_i-\bbox{\rho}_j)>$.
This   is  the  well known 
Lorentz-Lorenz local field i.e. the dipolar field at $\bbox{\rho}_i$ 
due to dipoles  at positions $\bbox{\rho}_j$ averaged over the 
position of the dipoles. 
Using well known results from the theory of dielectrics\cite{kit66} we obtain 
the second term in Eq.~(\ref{eq:moti}). Either 
for a random isotropic distribution of
 dipoles or for a cubic array  $L=1/3$. For other distributions $L$ can
 be computed using the results  of Ref.~\cite{mue35} (see below).

By putting  $\bbox{E}=\bbox{E}_0  e^{-i\omega t}$ we solve for 
the dipole moment $\bbox{\Pi}$ and
compute the polarization vector $\bbox{P}\equiv n \bbox{\Pi}$ where in the original 
variables the dipole moment is 
$\bbox{\Pi}_i\equiv -e \bbox{r}_i + q \bbox{R}_i$ .

 Finally from the relation between $\bbox{P}$  and $\bbox{E}$  one 
obtains the following dielectric function for  the interacting polaron liquid
\begin{equation}
  \label{eq:epsint}
  \epsilon(\omega)=1 - 
\frac{\Omega_p^{2}}{\omega^2+i \gamma \omega}
-   \frac{\omega_p^{2}-\Omega_p^{2}}{\omega^2-
\omega_{\rm coll}^2+ i\gamma' \omega}
\end{equation}
where we have introduced phenomenological inverse relaxation times 
$\gamma,\gamma'$ and defined the  bare plasma frequency 
$\omega_p^{2}\equiv 4\pi n e^2 /m$ and the polaron plasma frequency 
$\Omega_p^{2}\equiv 4\pi n e(e-q) / (M+m)=4\pi n e^2 /\epsilon_0 (M+m)$. In
the latter $M+m$ can be identified with the polaron  effective mass. 
 We have also
 defined the renormalized frequency of the internal mode
\begin{equation}
  \label{eq:ep}
  \omega_{\rm coll}^2\equiv\omega_0^2-\frac{q}{e}L (1+\frac{m}M)\omega_p^{2}
\end{equation}
where $\omega_0\equiv\sqrt{ k/\mu}$ is the frequency of the internal mode of
an isolated polaron.

The dielectric function Eq.~(\ref{eq:epsint}) takes a very simple form:
 the sum of a Drude term plus a collective excitation at frequency  
$\omega_{\rm coll}$. The latter  
consists of all  electrons oscillating in phase inside their respective
self-trapping potentials. In this oscillation  polarons develop 
time-dependent in-phase dipole moments. The dipole-dipole interaction
tends to soften this mode  and the effect increases as 
interparticle distance decreases i.e.  density increases 
($\propto \omega_p^2$).\cite{note4}  
Eq~(\ref{eq:epsint}) can be seen as a generalization of both Drude model and   
Clausius-Mossotti equation\cite{kit66}. Notice that  the f-sum rule is 
exactly satisfied. 

 At some critical density given by 
$4\pi n_c/e^2 m =\omega_0^2 e/q L(1+m/M)$
the energy cost to displace the electrons away from the respective
lattice distortion vanishes and an instability occurs.
 Within this
simplified model we cannot describe the new phase that arises for
$n>n_c$. In fact we assumed above that  electron and  distortion are bound 
and, even more, that the binding potential is harmonic. Both approximations 
will break down close to $n_c$. Interestingly, 
if a positive quartic term were present in the distortion-electron potential, 
 the system would become a liquid crystal of ferroelectric polarons above the 
critical density.

 A less  exotic possibility is that polarons start to dissociate.
This can occur either abruptly or in a continuous way. In the former case 
all polarons collapse at $n_c$ whereas  the latter case can be realized 
by a  two component system of coexisting polarons and free 
electrons\cite{ala90,lor94,ran97,fra98}.

Recently the softening of a polaron band has been observed as a function of 
doping\cite{lup99} in the Nd$_2$CuO$_{4-y}$ system.
We have used Eqs.~(\ref{eq:epsint}),(\ref{eq:ep})  to analyze those data. 

\begin{figure}[htbp]
   \epsfverbosetrue
\epsfxsize=6cm
$$
\epsfbox{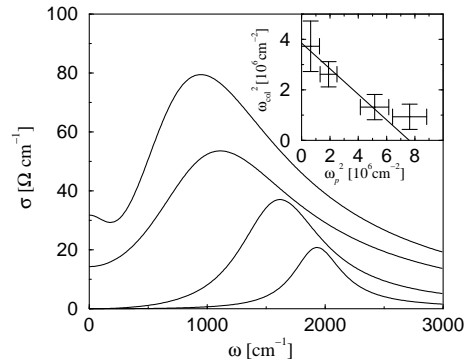}
$$
   \caption{Optical conductivity obtained from the fit of the model dielectric 
function to  reflectivity\protect\cite{lup92,cal94,lup99} data at 300K as 
explained in the text. Only Drude and internal mode polaron contributions 
have been included.  The upper curves have 
larger doping $y$ or equivalently larger density of particles $n$.  
The area under the curves is proportional to $\omega_p^2$. 
The maximum of the broad structure approximately corresponds to 
$\omega_{\rm coll}$.
Inset: The parameter $\omega_{\rm coll}^2$ vs. $\omega_p^2$ from the fits.
The error bars reflect  uncertainties in the fit. The line is a linear 
regression excluding the higher doping point.  }
    \label{fig:epdwp}
 \end{figure}

Slightly doped cuprates show a polaronic band, phonon bands, 
a mid-IR band and a charge transfer band\cite{lup92,cal94,tho91,fal93} 
(we neglect much weaker magnetic bands\cite{per93,lor95}). 
 Following Calvani and collaborators
 we fitted  reflectivity data\cite{lup92,cal94,lup99} on  
Nd$_2$CuO$_{4-y}$ with a
 Drude-Lorentz model for the dielectric constant (a sum of 
Lorentzians plus a Drude term, see Ref.~\cite{lup92}). 
The model dielectric function is of the same form as in 
Eq.~(\ref{eq:epsint})  but for the addition of phonons and
of high energy electronic contributions, and with an appropriate 
$\epsilon_{\infty}$ replacing 1. 
Because of the  electronic screening $\omega_p$ and $\Omega_p$  
 in  Eq.~(\ref{eq:epsint}) should be  considered 
screened plasma frequencies.
Two Lorentzians
were used above $10^4$ cm$^{-1}$ to fit the charge transfer band and
higher energy contributions.  Four Lorentzians were used below 800  cm$^{-1}$ 
to model the TO phonons, one Lorentzian above ~ 5000 cm$^{-1}$ (depending 
on  sample) were used to fit the mid-IR band (MIR) and finally one
Lorentzian below 2000 cm$^{-1}$  were used to fit the internal mode of 
polarons.
In Fig.~(\ref{fig:epdwp}) we show the resulting 
optical conductivity $\sigma(\omega) \equiv {\rm Im}\epsilon(\omega) \omega/4 \pi$, 
excluding the electronic contributions and the phonons.
 Neither $n$ nor $y$ are well  controlled 
variables so we use $\omega_p^2 (\propto n)$  as our control 
parameter which can be determined directly by adding the polaron and the Drude
oscillator strength [Eq.~(\ref{eq:epsint})]. In the inset of Fig.~\ref{fig:epdwp} 
we show $\omega_{\rm coll}^2$ vs.  $\omega_p^2$. Both quantities are obtained 
from the fits.

As found by Lupi {\it et al.}\cite{lup99} in this doping range
 we see a rapid decrease of the
 internal mode energy. To estimate the rate of decrease 
 we neglect the higher doping point (it is in the region where our 
approximations break down) and we do a linear regression. We obtain a slope of 
 $-0.5$.

To get a theoretical estimate of the slope in Eq.~(\ref{eq:ep}) 
 we need the  Lorentz-Lorenz local field  factor $L$. Using the
results of Ref.~\cite{mue35} we estimate $L=.6$ for Nd$_2$CuO$_{4-y}$
and for the electric field of the radiation parallel to the Cu-O plane.

Also the induced charge in the 
{\em lattice} will not be given any more by Eq.~(\ref{eq:q}). 
In fact  Eq.~(\ref{eq:q}) gives the total induced charge but part of this 
charge which we call $q_e$ is induced in the electronic degrees of freedom. 
This should be  subtracted from $+q$ and added to $-e$ since it will follow
 antiadiabatically the electron and hence is positioned at
 $\bbox{r}_i$, not at $\bbox{R}_i$. This effect is taken into account by  
 replacing $q/e$ by $(q-q_e)/(e-q_e)$ in Eq.~(\ref{eq:ep}).
 $q_e$ can be estimated,  using long wave length Lyddane-Sachs-Teller like 
arguments\cite{kit66}, to be of the order of  
$q_e= e ( 1-1/{\epsilon_\infty})({1+2/\epsilon_0})/({1+2/\epsilon_\infty})$.
  Using $\epsilon_0=30$ and  $\epsilon_{\infty}=5$\cite{lup92} 
one gets $(q-q_e)/(e-q_e)=0.91$.  Finally, as a first approximation,
we can take $(1+M/m)\sim 1$  so we obtain for the slope -0.55 in excellent 
agreement with the experiment. The  linear regression  extrapolates to 
$\omega_{\rm coll}=0$ at a critical value 
$\omega_{p,c}^2=7.6\times 10^6$ cm$^{-2}$.

For  $n>n_c$ a sudden increase of  Drude weight is observed (not shown), 
that we associate with metallization. This occur at much lower doping than 
the insulator superconducting transition observed at lower 
temperatures\cite{fou98}. 
We believe that more complicated phenomena like charge ordering may change 
the picture at low temperatures\cite{lup99}. Quantum corrections will also be
important close to $n_c$ and at low temperatures. 

 Lupi {et al.} find that the  collective mode subsists
at a small finite frequency for $n>n_c$ and then  monotonously decreases as
doping is increased at a much slower rate\cite{lup99}. 
This suggests a scenario where the polarons stabilize at a concentration
slightly below $n_c$ and added carriers go to free states. Different 
versions of such two-fluid model has been considered  in the 
literature\cite{ala90,lor94,ran97,fra98}.
Interestingly, this  coexistence implies that for a finite doping range
the system is at the verge of a dielectric anomaly with a soft
electronic mode. At larger doping, polarons
are expected to become unstable due to short-range interactions and screening
and a first order transition will occur to a one component Fermi liquid. 
We speculate that this scenario can explain the phase diagram of
cuprates.  In fact the soft collective mode can explain many of the anomalous 
Fermi liquid properties\cite{note2} that these systems show at optimum doping 
levels 
including pairing formation and superconductivity due to  exchange of 
the soft mode boson. At overdoping the experimental optical conductivity 
shows a
single component Drude behavior which  suggest a first order
transition\cite{uch91}. A related scenario including the softening of
an optical electronic mode associated with matallization has been
proposed previously\cite{lor93a}.

 The present theory also explains why a material like doped La$_2$NiO$_4$ 
never becomes
metallic. In this system the observed energy of the internal mode for small 
doping is  roughly by a factor of 3 larger than in  
Nd$_2$CuO$_{4-y}$ \cite{xia93,xia93b}. Then  if we 
assume that the slopes of $\omega_{\rm coll}^2$ vs.  $\omega_p^2$  are 
similar in the 
two systems  we expect, according to Eq.~(\ref{eq:ep}), that the critical 
plasma 
frequency $\omega_{p,c}^2$ in the nickelate  is by a factor of 9 larger
than in the cuprate. On the other hand, the observed spectral weight in this 
system reaches a maximum and decreases again as a function of 
doping\cite{xia93,xia93b}. We estimate the maximum $\omega_p^2$ by
integrating the fitted line shapes in Ref.~\cite{xia93,xia93b} to be 
roughly $9 \times 10^{6}$cm$^{-2}$ which is almost by an order of magnitude 
smaller than the estimated  $\omega_{p,c}^2$. This means 
that on undimensionalized scales 
($\omega_{\rm coll}^2/\omega_0^2$ vs. $\omega_p^2/\omega_{p,c}^2$) 
only the first $\sim 10\%$ of the curve  in the inset of 
Fig.~\ref{fig:epdwp} is 
physically accessible and the system {\em never reaches the point 
in which it should metallize}. 
Meaningfully  a modest softening of the
internal mode {\em is  actually observed}\cite{xia93,xia93b} in this  
relatively small $\omega_p^2/\omega_{p,c}^2$ variation range consistent  
with Eq.~(\ref{eq:ep}). 

To conclude we have derived a many body Hamiltonian that takes into account 
dipole-dipole interactions in a polaron system in a general way. We 
obtained the dielectric constant in mean field for a liquid of such 
particles and showed that the
system becomes unstable for large concentration. We argue
that the theory explains a rapid softening of the internal mode of the
polaron recently observed in cuprates\cite{lup99} and discussed the 
 doping-induced matallization of cuprates and nickelates. We speculate
 that the same framework can be used to explain the occurrence of
 matallization or not as a function of doping in a wide range of 
 materials where the electron phonon interaction dominates.  

We thanks P. Calvani, S. Lupi, M. Capizzi, S. Fratini and
P. Qu\'emerais  for enlightening discussions and Lupi {\it et al.} 
for providing us their  reflectivity data before publication.


\begin{references}
\bibitem{note4}
The instability is a very general phenomena and will occur even if the polarons
  are localised either randomly or in a crystalline lattice. A closely
related instability and optical conductivity in
  a Wigner crystal of polarons has been obtain in Ref.~\cite{fra98}.
\bibitem{note2}
Non Fermi-liquid behavior has been found due to coexistence of localized states
  and free electrons in Ref.~\cite{ala90}. This model however does not invoke a
  soft electronic mode in the sense we discuss.
\end{references}

\end{document}